\documentclass[twocolumn,showpacs,amsmath,amssymb]{revtex4}

\usepackage{graphicx}
\usepackage{epsfig}
\usepackage{psfrag}
\usepackage{dcolumn}
\usepackage{bm}

\def\kp{${\bf k}\cdot{\bf p}$}

\begin{document}
\title{Spin-Orbit Effects in Diluted Magnetic Semiconductors}

\author{Byounghak Lee$^1$, Xavier Cartoix\`a$^1$, Nandini Trivedi$^2$, and Richard M. Martin$^1$}
\affiliation{$^1$
Loomis Laboratory of Physics, University of Illinois at Urbana-Champaign, 1110 W. Green Street, Urbana, Illinois 61801
}
\affiliation{$^2$
Department of Theoretical Physics, Tata Institute of Fundamental Research, Homi Bhabha Road, Mumbai 400005 INDIA
}

\date{\today}

\begin{abstract}
We present a theoretical study of diluted magnetic semiconductors that
treats the local {\it sp}-{\it d} exchange interaction $J$
between the itinerant carriers and the Mn {\it d} electrons
within a realistic band structure and goes beyond previous mean-field approaches.
In case of Ga$_{1-x}$Mn$_x$As, we find that strong exchange potentials tightly bind carriers near impurity sites.
Spin-orbit coupling is found to have a more pronounced effect on spin polarization at weak coupling
and this feature can be exploited for magnetotransport.
For low doping regime, we predict that spin-orbit coupling splits impurity bands and can induce a novel insulating
phase.

\end{abstract}


\maketitle

It has already been more than half a decade since ferromagnetism with Curie
temperatures T$_c \sim$ 100 K was discovered in III-V diluted magnetic semiconductors (DMS)~\cite{Ohno}, and following that a large amount of experimental knowledge
has been accumulated~\cite{Awschalom book}.
There also has been a great deal of theoretical activity~\cite{Byounghak 1999, Konig and MacDonald, Schliemann and MacDonald, Dietl and Ohno, Berciu and Bhatt, Das Sarma and Millis, Alvarez and Dagotto, Timm et al, Tang and Flatte 2004} to understand the  mechanism of ferromagnetism, but consensus has not yet been reached.
The generally accepted theoretical picture is that
the long range ferromagnetic order between the magnetic impurities with large magnetic moment is mediated via  the interaction with the itinerant carriers~\cite{MacDonald review}.
The comprehensive theoretical understanding, however, has been hampered by the complexity of the system arising from the interplay of equally important but distinct effects, namely the electron-local moment interaction, disorder effects, and spin-orbit (SO) effects.

Transport properties~\cite{van Esch et al} indicate that DMS do not belong to either of the well established classes of magnetism, i.e., itinerant ferromagnetism and local moment ferromagnetism, but rather are intermediate in character.
Disorder plays the major role in determining the nature of the metal-insulator transition~\cite{Timm et al, Berciu and Bhatt}.
In addition, in typical DMS, the orbital degrees of freedom couple to the spin degrees of freedom via relativistic effects.
In III-V semiconductors, the SO coupling ranges from a few tens to several
hundreds of meV, in particular the splitting of the bands at {\bf k}=0 is 0.34 eV for GaAs~\cite{Vurgaftman and Meyer 2001}.
This energy is as big as the Fermi energy in the relevant impurity doping concentrations.
The approaches based on itinerant ferromagnetism often assume a continuum distribution of magnetic impurities interacting with host valence band electrons~\cite{Byounghak 1999, Konig and MacDonald, Dietl and Ohno}.
While this gives good descriptions in the metallic regime, it is not applicable when the local character of the magnetic interaction is important, not to mention the disorder effects that are ignored completely.
More realistic tight-binding models based on atomic orbitals have been limited to the dilute impurity regime~\cite{Tang and Flatte 2004}.
The alternative approach of local moment magnetism has neglected spin-orbit coupling and
used only a  single band for each spin species~\cite{Berciu and Bhatt, Das Sarma and Millis, Alvarez and Dagotto},
largely due to the computational complexity.
This, however, cannot be justified without an obvious symmetry breaking which completely lifts the orbital degeneracy.
Density functional theory (DFT) calculations~\cite{Sanvito and Hill 2001, dft calculations} suffer high computational cost in dealing with systems with dilute impurities, and more complicated issues such as disorder effects are hard to study using small systems.
The purpose of this paper is to introduce a robust model which can be applied to wide ranges of material parameters without ignoring any of the important effects mentioned above, and, as an immediate example, to explore the SO effects on the ground state electronic structures.

Our main results are as follows:
(1) The binding potential of magnetic impurity is strong enough to generate significant weight of the carriers of one spin at the impurity site.
This results in a large splitting between the majority and minority spin bands, and leads to qualitatively different band structures from continuum model predictions.
(2) The spin polarization of itinerant carriers increases with the interaction with magnetic impurity, $J$, for fixed doping concentration, $x$, and also increases with $x$ for a fixed $J$.
The intrinsic spin-orbit coupling effects reduce the degree of spin polarization and have relatively bigger effects on spin polarization at low $x$.
The ability of manipulate the spin polarization by doping is attractive for magnetotransport applications.
(3) For low doping, spin orbit and exchange couplings lift the degeneracies of the impurity
bands resulting in an insulator with the chemical potential in the band gap.

We describe DMS following the $sp$-$d$ exchange model~\cite{Kossut}
containing the Hamiltonian of the host semiconductor in the clean limit and the effective exchange coupling between the localized and itinerant spins.
Since the relevant energy scale is only order of a few hundred meV, it is of crucial importance to use a host semiconductor Hamiltonian which is precise close to the Brillouin zone center.
To this end, we use the Effective Bond Orbital Model (EBOM) developed by Chang~\cite{Chang 1988}.
EBOM has previously been applied to DMS within a continuum mean-field approximation~\cite{Vurgaftman and Meyer}.
The difference in our model is the explicit inclusion of the localized character of the exchange interaction.
Our basis set is composed of the bulk state solutions at {\bf k}=0, which are Bloch sum of effective bond orbitals located at fcc lattice sites.
\begin{figure}[t]
  \centerline{\includegraphics[width=0.7\columnwidth]{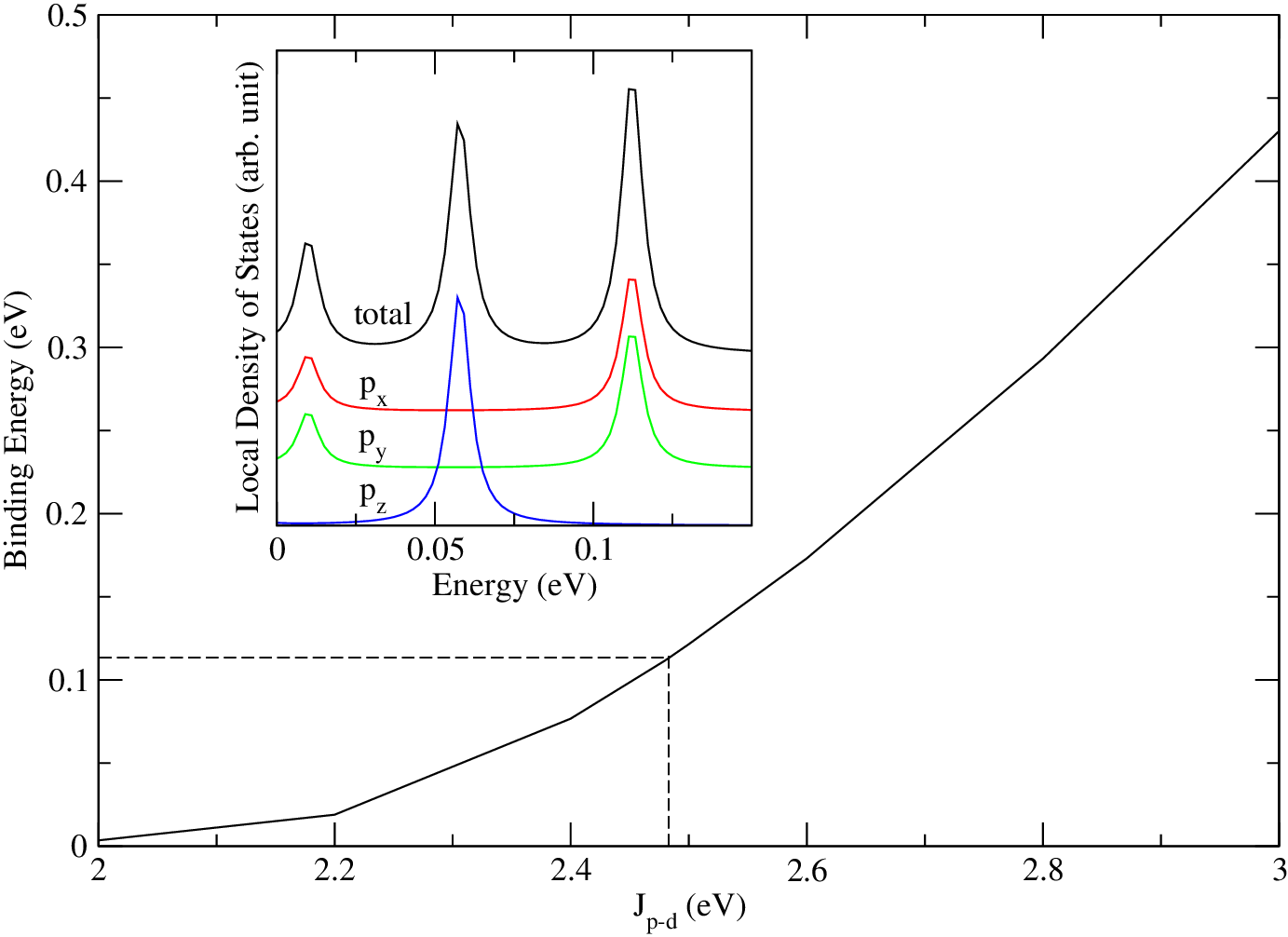}}
    \caption{
      The impurity binding energy as a function of exchange coupling constant $J_{p-d}$.
      The experimental value is indicated by the dashed line.
      The inset shows the orbital-resolved local density of states at the impurity site with the fitted exchange coupling of J$_{p-d}$=-2.48 eV.
      The density of states are vertically shifted for clarity.
    }
    \label{fig:ImpurityStates}
\end{figure}
This basis set has the nice feature that the second order expansion of the matrix elements about the zone center yields the results of the \kp\ model.
In addition, it reproduces the experimental band structure over wider ranges of the Brillouin zone.
The impurities are placed at the fcc lattice sites and the matrix elements are calculated in the effective bond basis.
The applicability of this scheme in calculating impurity states has been well demonstrated~\cite{Einevoll and Chang}.
The exchange couplings are responsible for the hybridization of impurity states with itinerant electron states as well as the interaction between electrons at impurity sites.
In general, it is possible that the contribution from electron-electron interaction can give rise to a dependency of exchange coupling constants on magnetic impurity concentrations.
We believe, however, this dependency is negligible because of the large dielectric constant of GaAs and the long Fermi wavelength of typical DMS systems.
The resulting Hamiltonian can be written as
\begin{eqnarray}
H &=& \sum_{i,j,n,m} t_{i j}^{n m} c_{i n}^{\dagger} c_{j m} + \sum_{i,n,m} E_i^{n m} c_{i n}^{\dagger} c_{i m} \nonumber \\
& & - \sum_{I,n,m} J_{n m} {\bf S}_I c_{I n}^{\dagger} {\bm \sigma}_{n m} c_{I m} \;,
\label{eq:Hamiltonian}
\end{eqnarray}
where $i$, and $j$ are the fcc site indices,
$I$ is the location of the substitutional impurities on the fcc sites,
${\bm \sigma}$ is the Pauli spin matrix, and ${\bf S}_I$ is the local spin moment, which we treat as a classical object of magnitude $S$=5/2.
$n$ and $m$ are the combined indices of the spin and the bond orbital, i.e., $n=\{\tau, \alpha\}$, where in the current 8-band model $\tau$=$\{\uparrow, \downarrow\}$ and $\alpha$=$\{s,p_x,p_y,p_z\}$.
The first two terms on the right hand side of (\ref{eq:Hamiltonian}) are the hopping and on-site potential of EBOM, respectively.
The SO coupling is included in the onsite matrix elements $E_i^{nm}$.
The last term in Eq.~(\ref{eq:Hamiltonian}) is the exchange interaction $H_{ex}$, whose coupling matrix $J_{nm} = -\langle n | H_{ex} | m \rangle(2/S)$ is to be determined as discussed below.
Our formalism is general enough to include disorder~\cite{unpublished}; however in this paper we calculate
the electron states for supercells made of fcc lattices with a single magnetic impurity oriented in the z-direction (say).
The magnetic impurity concentration, $x$, is defined as the inverse of the number of fcc sites.

The exact value of the exchange coupling is rather unsettled.
Magneto-reflectivity~\cite{Szczytko et al 1996}, core-level photoemission~\cite{Okabayashi et al 1998}, resonant photoemission spectroscopy~\cite{Okabayashi et al 1999}, and magneto-transport~\cite{Matsukura et al 1998} measurements found 2.5 eV, -1.0 eV, -1.2 eV, -3.3 eV for the p-d exchange coupling, respectively.
To circumvent this ambiguity, we determine the exchange coupling matrix by comparing the impurity bound state energy with experimental data.
Because Mn atoms in GaAs are substitutional impurities with the site symmetry of the Ga sites~\cite{Shioda et al 1998}, it can be readily shown that the exchange coupling does not mix different orbital states.
The exchange coupling of conduction electrons with Mn 3d electrons, $J_{s-d} \equiv J_{\tau s, \tau s}$, are known to be very weak, generally of order of 0.1 eV~\cite{Szczytko et al 1996}.
We found that our results are very insensitive to this coupling.
Hereafter, we set the coupling to the conduction electrons to 0.1 eV.

\begin{figure}[t]
  \centerline{
   \includegraphics[width=0.8\columnwidth]{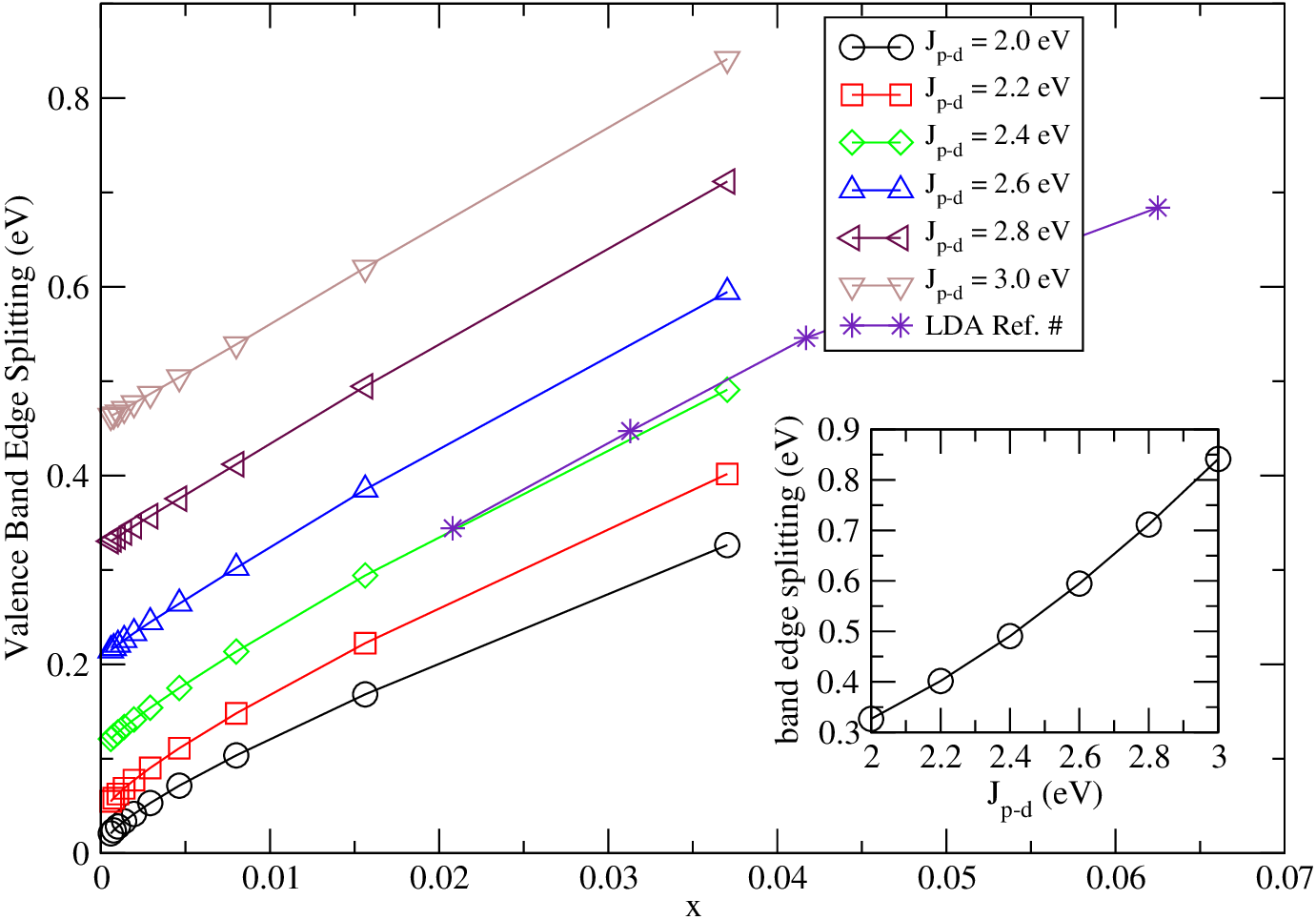}}
    \caption{
      The valence band-edge splitting in the absence of spin-orbit coupling as a function of impurity concentration for various exchange couplings $J_{p-d}$. The star symbols are the data from Ref.~\cite{Sanvito and Hill 2001}.
      The inset shows the valence band-edge splitting as a function of exchange coupling at a fixed impurity concentration of $x$ = 0.037.
    }
    \label{fig:band_splitting_woLS}
\end{figure}

In Fig.~\ref{fig:ImpurityStates} we show the binding energy of a single impurity as a function of the exchange coupling.
The bound state energies were extrapolated from supercell calculations of $13^3$ to $15^3$ fcc lattice systems.
With the best fitted exchange coupling $J_{p-d}$ = -2.48 eV, the impurity bound state energy agrees with experimental
observation~\cite{Lee and Anderson 1964} of 0.113 eV.
The orbital-resolved local density of states, presented in the inset of Fig.~\ref{fig:ImpurityStates}, shows that the bound states are in almost definite spin states and split into orbital angular momentum substates of m$_z$=-1, 0, and 1.
This agrees with recent calculation by Tang and Flatt\'e based on atomic orbital tight-binding Green's function method~\cite{Tang and Flatte 2004}.
It is worth noting that there is a critical magnitude of the exchange coupling
$\sim$ 1.8 eV, below which the valence holes are not bound to impurities.

\begin{figure}[t]
  \centerline{ \includegraphics[width=0.8\columnwidth]{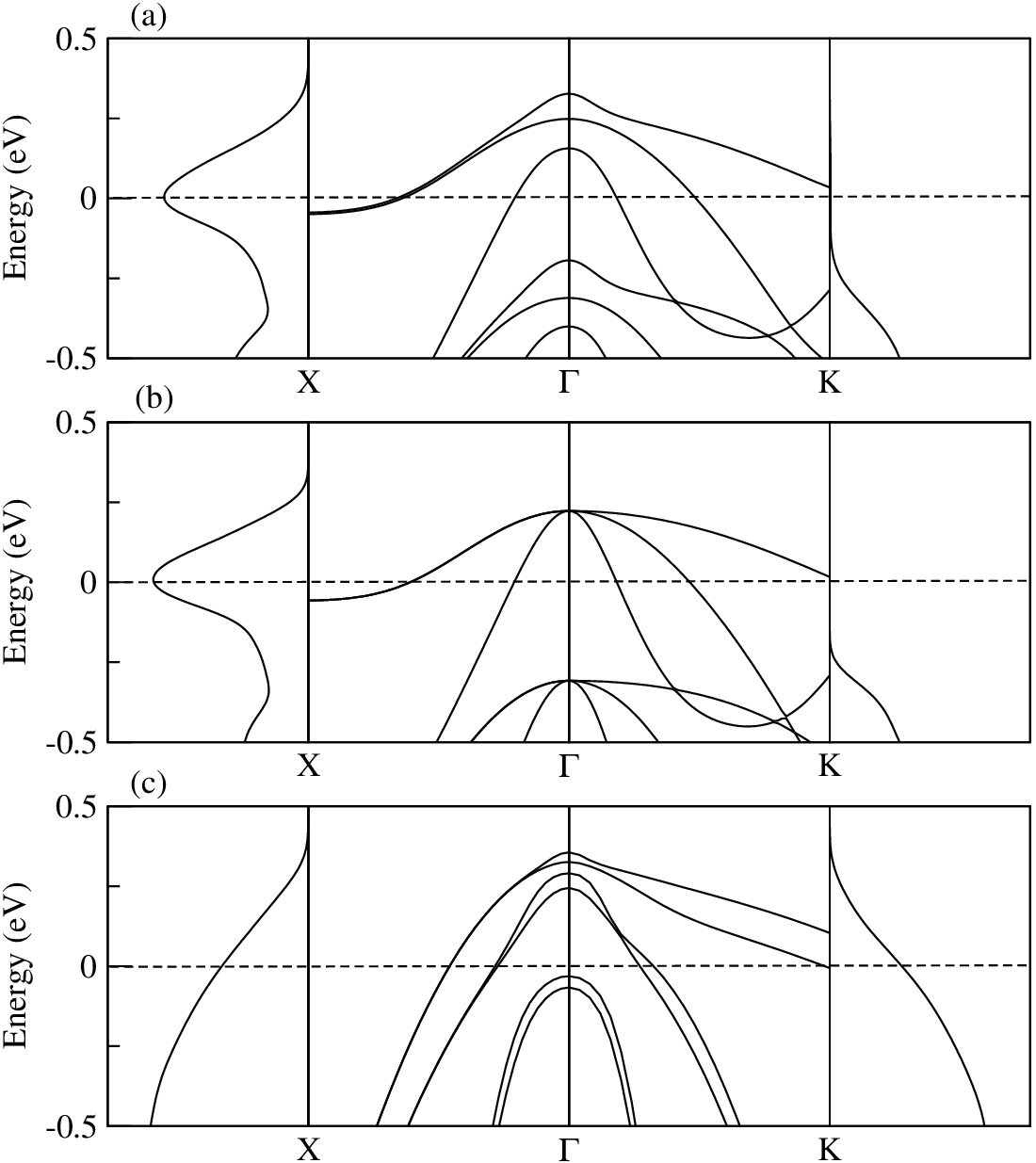} }
  \caption{
    (a) The valence band structures of Ga$_{0.963}$Mn$_{0.037}$As calculated from EBOM model with exchange coupling $J_{p-d}$ = -2.48 eV.
    The Fermi level is depicted by the horizontal line.
    The spin-resolved density of states is drawn on the left (right) for majority (minority) spins.
    (b) Same figure as (a) except calculated in the absence of SO coupling.
    (c) \kp\ band structure with mean-field exchange coupling $\beta N_0$ = -1.2 eV and $x$ = 0.037.
  }
  \label{fig:metal_bands}
\end{figure}

The applicability of our model to dense impurity concentration regime is supported by a comparison with existing {\em ab initio} DFT calculations~\cite{Sanvito and Hill 2001, dft calculations}.
In the DFT calculation by Sanvito {\it et al.}~\cite{Sanvito and Hill 2001}, the band-edge splitting of Ga$_{1-x}$Mn$_x$As was calculated as a function of the impurity concentration.
We performed equivalent supercell calculations using our model Hamiltonian, turning off the SO coupling for comparison purposes.
In Fig.~\ref{fig:band_splitting_woLS} we present the band-edge splitting of the majority and minority valence bands.
At $J_{p-d} \equiv J_{\tau p_i, \tau p_i}$ = -2.40 eV, our calculation shows an excellent agreement with the existing DFT calculation in the region $x$ $\sim$ (0.01, 0.04)~\cite{Sanvito and Hill 2001}.
The fitted exchange coupling differs only by 0.08 eV from the fitting to the single impurity bound states.
This is an encouraging result since it confirms that our model is applicable in a wide range of impurity concentrations.
The calculated band structure with the fitted coupling parameter, Fig.~\ref{fig:metal_bands}(b), also agrees very well with previous DFT calculations~\cite{Sanvito and Hill 2001, dft calculations}.
The nonlinear dependency of valence band-edge splitting on the exchange
coupling, inset of Fig.~\ref{fig:band_splitting_woLS}, shows a clear
difference between our model and the virtual crystal approximation (VCA), which predicts a linear relation when SO coupling is not present.

Having established the model, we turn to assessing the effects of SO coupling on the electronic structure.
Most of DFT calculations have predicted half-metallicity in Ga$_{1-x}$Mn$_x$As~\cite{Sanvito and Hill 2001, dft calculations} and this prediction has drawn much research interest in regards to possible application to the spin devices~\cite{Awschalom book}.
When the spin degree of freedom is separated from orbital degrees of freedom, it is natural to have spin-split bands in the presence of magnetic background.
In Fig.~\ref{fig:metal_bands} (b) we show the band structure of Ga$_{0.963}$Mn$_{0.037}$As when SO coupling is not considered.
The separation of orbitally degenerate majority and minority spin bands agrees very well with DFT calculations~\cite{Sanvito and Hill 2001}.
On the contrary, in \kp\ approximations, where the impurities are treated within a virtual crystal approximation, the exchange interaction is scaled by impurity concentration and ends up introducing a relatively weak potential.
Fig.~\ref{fig:metal_bands} (c) shows a typical \kp\ bands with mean-field exchange coupling $\beta N_0$ = -1.2 eV~\cite{Dietl and Ohno}.
The exchange coupling splits the heavy- and light-hole bands as little as 0.1 eV and the mixing of light- and heavy-hole bands leads to weak spin polarization.
Our calculation resolves the conflict between above two predictions.
Taking into account both SO coupling and local exchange interaction, we found that electron states of Ga$_{1-x}$Mn$_x$As with dense impurity concentration, i.e. $x$ $>$ 0.01, is indeed half-metallic.
Fig.~\ref{fig:metal_bands} (a) illustrates that the splitting due to exchange coupling is much larger than the SO coupling and the low states are fully spin polarized.

\begin{figure}[t]
\centerline{ \includegraphics[width=0.75\columnwidth]{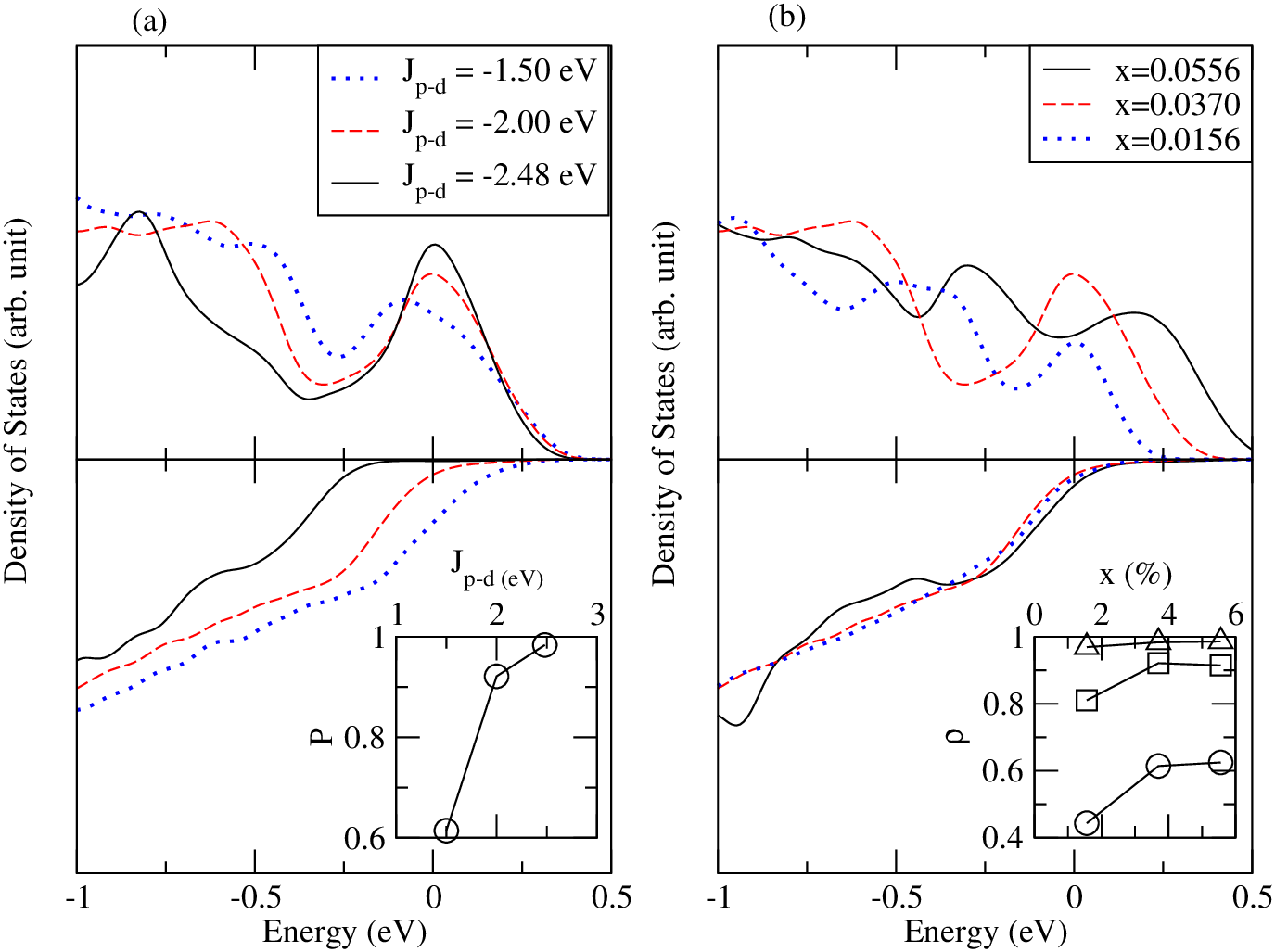} }
\caption{
Density of states Ga$_{1-x}$Mn$_x$As.
(a) Magnetic impurity concentration $x$ = 0.037. Inset shows the polarization as a function of exchange coupling.
(b) Exchange coupling, $J_{p-d}$ = -2.0 eV. Inset shows the polarization as a function of impurity concentration. Circle, square, and triangle correspond to $J_{p-d}$=-1.5 eV, $J_{p-d}$=-2.0 eV, and $J_{p-d}$=-2.48 eV, respectively.
For both (a) and (b) the upper panel is the majority spin density of states and the lower panel is the minority spin density of states.
The energy is relative to the Fermi energy located at zero.
}
\label{fig:SpinSplit}
\end{figure}

To explore the dependency of the spin-polarization, we present the spin-resolved density of states for various exchange couplings and impurity concentrations in Fig.~\ref{fig:SpinSplit}.
When exchange coupling is not strong enough to induce impurity bound states, e.g. $J_{p-d}$ $\sim$ -1.5 eV, the minority spin has substantial contribution to the polarization.
As the exchange coupling gets stronger the polarization becomes bigger, reaching near 100\% at $J_{p-d}$ $\sim$ -2.48 eV.
The dependency of spin polarization on the impurity concentration is of more practical interest since it can be controlled in sample growing processes.
The increase in the impurity concentration contributes to the spin-polarization by enhancing the splitting between the majority and minority bands.
We found the spin polarization monotonically increases within the experimentally accessible impurity concentration range $x$ $\le$ 0.1.
%\vspace{.5cm}

\begin{figure}[t]
  \centerline{ \includegraphics[width=0.7\columnwidth]{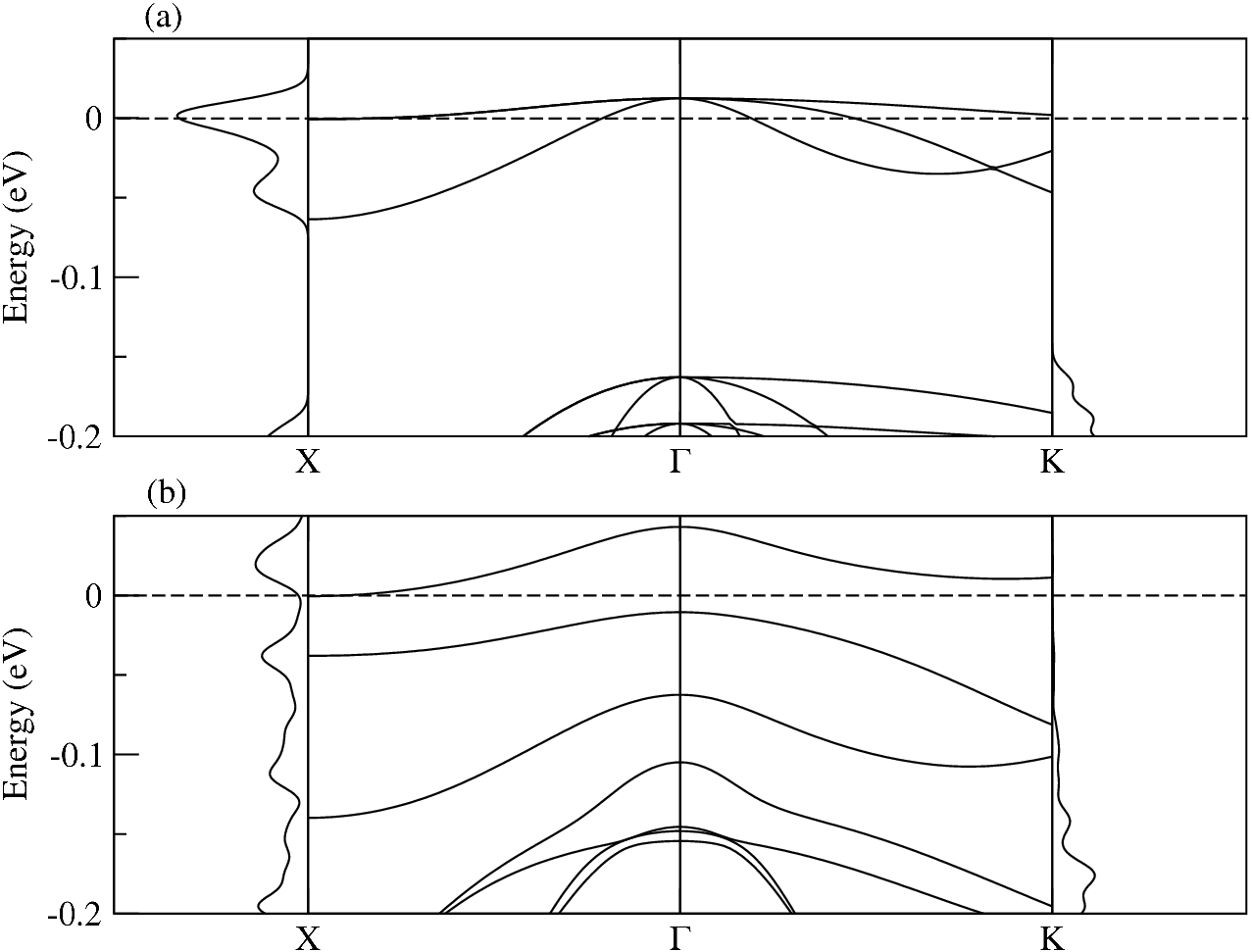} }
\caption{
The valence band structures of Ga$_{0.998}$Mn$_{0.002}$As.
(a) without spin-orbit SO coupling. (b) with SO coupling. The dotted line shows the location of the Fermi level.
Note that in (b) the Fermi level lies in the band gap generated largely by SO effects.
}
\label{fig:impurity_band}
\end{figure}
At low impurity concentrations, the SO coupling can be comparable to the width of the impurity
bands and this can result in a novel type of insulating phase.
The mechanism by which such an insulator is formed is discussed below.
\noindent
A single orbital is split into two spin polarized bands by the exchange interaction.
In the absence of charge compensation the majority spin band is completely filled while the minority band is empty.
Since the chemical potential lies in the middle of the spin-split bands, the system is insulating.
In a real semiconductor, on the other hand, with several degenerate orbitals in the valence band,
the degeneracy can be lifted by two effects arising from the exchange coupling as well as from the spin-orbit coupling.
The exchange coupling acting alone splits the orbitals into partially occupied majority bands and empty minority bands
from which one would have concluded that the system is metallic as seen in Fig.~\ref{fig:impurity_band}(a).
However, the presence of spin-orbit coupling further splits the bands
and opens up a gap. For sufficiently narrow impurity bands, the chemical potential then lies
within the gap and results in an insulating state as seen in Fig.~\ref{fig:impurity_band}(b).
In strongly SO coupled systems, such as InSb, one might be able to find the insulating phase driven by the SO coupling even with moderate disorder effects.
%The disorder effects and the electron-electron interactions can be included within our model.
Disorder not only broadens the impurity bands but also reduces the band splitting, resulting in the merge of impurity bands into the main valence bands in case of Ga$_{1-x}$Mn$_x$As~\cite{unpublished}.

In summary we presented a tight-binding model which properly includes SO coupling effects in Ga$_{1-x}$Mn$_x$As.
Our model reproduces the experimentally observed impurity binding energy in dilute impurity limit and agrees well with first principle calculations in dense impurity limit.
We found the impurity bound states are split due to the SO coupling.
In dense doping regime, the spin-polarization increases with magnetic impurity concentration and almost fully spin-polarized metallic states can be achieved.
We observe impurity bands formation at low impurity concentrations and predict that strong SO coupling can induce a novel insulating phase.

B.L. is grateful to Y.-C.~Chang for helpful discussions about EBOM.
B.L. and R.M.M. were supported by the Office of Naval Research under Grant No. N0014-01-1-1062.

\end{document}